\documentclass[twocolumn,prl]{revtex4}
\usepackage{graphicx}
\usepackage{amsmath}
\usepackage{amssymb}
\usepackage{bm}

\begin{document}
%
\title{Non-quantized Dirac monopoles and strings \\
in the Berry phase of anisotropic spin systems}

\author{Patrick Bruno}\email{bruno@mpi-halle.de}
\affiliation{Max-Planck-Institut f{\"u}r Mikrostrukturphysik,
Weinberg 2, D-06120 Halle, Germany} 

\date{April 26, 2004}

\begin{abstract}
The Berry phase of an anisotropic spin system that is
adiabatically rotated along a closed circuit $\mathcal{C}$ is
investigated. It is shown that the Berry phase consists of two
contributions: (i) a \emph{geometric} contribution which can be
interpreted as the flux through $\mathcal{C}$ of a
\emph{non-quantized} Dirac monopole, and (ii) a \emph{topological}
contribution which can be interpreted as the flux through
$\mathcal{C}$ of a Dirac string carrying a \emph{non-quantized}
flux, i.e., a spin analogue of the Aharonov-Bohm effect. Various
experimental consequences of this novel effect are discussed.
\end{abstract}

\maketitle
%
Dirac's elegant explanation of charge quantization as a
consequence of the existence of at least one hypothetical magnetic
monopole \cite{Dirac1931} has stimulated considerable interest. It
is important to realize that the hypothesis of spacial isotropy
plays a central role in Dirac's theory: spacial isotropy requires
that the Dirac string attached to the Dirac monopole (in order to
ensure compatibility with the known laws of electromagnetism and
quantum mechanics) be ``invisible" to electrons, which, in turn,
requires that the flux carried by the Dirac string is quantized in
units of $\phi_0\equiv hc/e$ (otherwise the string would be able,
via the Aharonov-Bohm effect \cite{Aharonov1959}, to scatter
electrons, which would violate the assumed isotropy). The
quantization of electric and magnetic charges in units of $e$ and
$g$, respectively, with $eg = \hbar /(2c)$, then follows. To the
best of our current experimental knowledge, space is indeed
isotropic and charges are quantized to a relative accuracy better
than $10^{-21}$ \cite{Baumann1988}. But in spite of considerable
efforts, Dirac monopoles have remained elusive, so far.

It is therefore of great interest to investigate objects that are
completely analogous to the ``real" Dirac monopoles, however
living in some abstract space (unlike the ``real" Dirac monopoles,
which live in real space), and possessing the great advantage of
being more easily amenable to experiment. Such ``fictitious" Dirac
monopoles (from here on, I shall simply call them Dirac monopoles,
without ambiguity) play an outstanding role in the context of the
Berry phase of quantum systems adiabatically driven around a
closed circuit $\mathcal{C}$ in the space of external parameters
\cite{Berry1984}. For the case of a spin $S$ in a magnetic field
(hereafter called Berry's model), the Berry phase is proportional
to the ``flux" of a \emph{quantized} ``Dirac monopole" through the
circuit $\mathcal{C}$ (in this case the external parameters reduce
to a unit vector and the parameter space is the sphere $S^2$). The
quantization of the ``Dirac monopole" is directly related to the
quantization of the angular momentum along the field axis, i.e.,
to the rotational invariance of the Hamiltonian around the field
axis. However, in the light of the above discussion on the
interplay between space isotropy and charge quantization, one can
anticipate that the quantization of the Dirac monopole may be
lifted if the rotational invariance is broken, which may lead to
non-trivial new physical phenomena. For Berry's model, the
parameter space (sphere $S^2$) is simply connected (its
fundamental homotopy group is trivial: $\pi_1(S^2)=0$), so that
the Berry phase may not depend on \emph{topological} properties of
the circuit $\mathcal{C}$ and is purely \emph{geometric} (solid
angle). By contrast, in the more general case of anisotropic spin
systems, the parameter space is not simply connected, as discussed
below, so that the Berry phase may be expected to contain a term
that depends on some \emph{topological} property of the circuit
$\mathcal{C}$. The simplest example of a \emph{topological} Berry
phase is given by the Aharonov-Bohm effect \cite{Aharonov1959}:
here the parameter space is (or may be reduced to) the circle
$S^1$, which is non-simply connected and has a non-trivial
fundamental homotopy group $\pi_1(S^1)=\mathbb{Z}$; the Berry
phase is given by the winding number of the circuit $\mathcal{C}$
around the Aharonov-Bohm flux tube multiplied by the (in general
\emph{non-quantized}) flux of the tube. The aim of the present
paper is (i) to show that something similar generally happens in
anisotropic spin systems, (ii) to give explicit predictions for
this novel effect, and (iii) to discuss some experimental
realizations.

Let $H_0$ be the Hamiltonian of a completely general spin system,
comprising an arbitrary number of interacting spins, subject to
external magnetic fields, and to arbitrary magnetic anisotropies,
and $H_{\mathbf{R}}\equiv U_{\mathbf{R}}H_0{U_{\mathbf{R}}}^{-1}$
the Hamiltonian resulting from a global rotation $\mathbf{R}$. We
are interested in the Berry phase associated with a closed circuit
$\mathcal{C}$ in the parameter space of rotations, consisting of
adiabatic continuous sequences of rotations $\mathbf{R}(t)$ with
$t \in [0,T]$ and $\mathbf{R}(0) = \mathbf{1}$. Obviously, for
$\mathcal{C}$ to be closed, $\mathbf{R}(T)$ has to belong to the
group $\mathcal{G}$ of the proper symmetries of $H_0$.

Let us first discuss the topology of the parameter space
$\mathcal{M}$. The latter depends on the order $q$ of
$\mathcal{G}$. As an example, let us consider the model depicted
on Fig.~1; for cases (a), (b), and (c), $\mathcal{G}$ is $C_1$,
$C_2$, and $C_\infty$, respectively. Rotations may be represented
in the axis-angle parameterization by points in a 3D-ball of
radius $2\pi$. Each rotation of $SO(3)$ is represented by a pair
of 2 distinct points in the ball (since $(\theta ,
\mathbf{\hat{n}})$ and $(2\pi-\theta , -\mathbf{\hat{n}})$
represent the same rotation); in particular, the identity
$\mathbf{1}$ is represented both by the origin and by the entire
sphere of radius $2\pi$. Furthermore, rotations of $SO(3)$ which
are related to each other by proper symmetries of $H_0$ yield in
fact the same Hamiltonian and are to be identified, so that each
element of the parameter space $\mathcal{M}$ is represented by a
set of $2q$ points in the ball of radius $2\pi$, i.e.,
$\mathcal{M}=SO(3)/\mathcal{G}$. For case (c)
($\mathcal{G}=C_\infty$), one has on the $z$-axis a continuous
line of points equivalent to the identity. Closed loops starting
from the origin that can be continuously deformed into each other
(shown with the same color in Fig.~1) belong to a same homotopy
class. For example one easily sees that there are, respectively,
2, 4, and 1 distinct homotopy classes in cases (a), (b), and (c).
To summarize, for $\mathcal{G}=C_q$ with $1\leq q < +\infty$, one
finds $\mathcal{M}=SO(3)/C_q = \mathbb{R}P^3/\mathbb{Z}_q$, and
the fundamental isotopy group is $\pi_1(M)=\pi_1
(\mathbb{R}P^3/\mathbb{Z}_q)=Z_{2q}$, the group of integers modulo
$2q$; for $\mathcal{G}=C_\infty$, one gets
$\mathcal{M}=SO(3)/C_\infty =S^2$, which is simply connected, and
one recovers the result of Berry's model,
$\pi_1(\mathcal{M})=\pi_1(S^2)=0$, as expected.

Let us then calculate the Berry phase. The rotations will now be
parameterized in the form $\mathbf{R}\equiv (\varphi,\theta,\beta
)$, with $\beta \equiv \varphi + \psi$, where
$(\varphi,\theta,\psi )$ are the Euler angles.
The polar angles $(\varphi,\theta)$ give the orientation of unit
vector $\hat{\mathbf{z}}(\mathbf{R})$ of the rotated $z$-axis,
while $\beta$ gives the twist angle of the $x$- and $y$-axes
around $\hat{\mathbf{z}}(\mathbf{R})$. The unitary operator of the
rotation $\mathbf{R}$ is
$ U_\mathbf{R} \equiv  \mathrm{e}^{ -\mathrm{i}\varphi J_z}
\mathrm{e}^{ -\mathrm{i}\theta J_y} \mathrm{e}^{ -\mathrm{i}(\beta
-\varphi ) J_z} $,
$\mathbf{J}$ being the total angular momentum operator.

\begin{figure}[t]
\includegraphics[width=1.0\columnwidth]{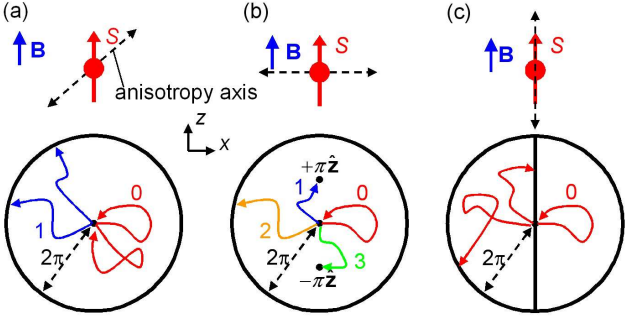}
\vspace*{-0.5cm} \caption{Sketch of the parameter space
$\mathcal{M}$ (in angle-axis representation) for a spin $S\geq 1$
with a uniaxial anisotropy axis of various orientations, and a
magnetic field along $z$. The various homotopy classes of closed
loops are shown and labelled by the corresponding element of the
fundamental homotopy group $\pi_1(\mathcal{M})$. See text for
details.} \label{fig_Spin}
\end{figure}

Let $\left| \Psi_0^{n} \right\rangle$ (with $n=1,2,\ldots$) be the
normalized eigenstates of $H_0$, of energies $E_n$. For a general
rotation $\mathbf{R}$, we choose as basis functions the rotated
eigenstates,
$ \left| \Psi_\mathbf{R}^{n} \right\rangle \equiv U_{\mathbf{R}}
\left| \Psi_0^{n} \right\rangle , $
with energies $E_n(\mathbf{R})=E_n$, independent of $\mathbf{R}$.
From now on, we shall restrain our discussion to the case of
non-degenerate levels $E_n$ (abelian case). Note that
$H_\mathbf{R}= H_0$ does \emph{not} imply $\left|
\Psi_\mathbf{R}^{n} \right\rangle = \left| \Psi_0^{n}
\right\rangle$ but only $\left| \Psi_\mathbf{R}^{n} \right\rangle
= \mathrm{e}^{-\mathrm{i}\alpha_\mathbf{R}^n} \left| \Psi_0^{n}
\right\rangle$; this multivaluedness of the basis functions
requires particular care.

Berry \cite{Berry1984} pointed out that for a system satisfying
$\left|\Psi (t=0)\right\rangle = \left| \Psi_0^{n} \right\rangle$
adiabatically (i.e., in a time $T\gg\hbar /|E_m - E_n|$, $\forall\
m\neq n$) transported along the closed circuit $\mathcal{C}$, the
wave function at time $T$ (after closing the circuit) is given by
$ \left| \Psi (T)\right\rangle =
\mathrm{e}^{\mathrm{i}\left[\delta_n +
\gamma_n(\mathcal{C})\right]} \left| \Psi_0^{n}\right\rangle, $
where
$ \delta_n \equiv -{\hbar}^{-1}\int_0^T E_n(\mathbf{R}(t))\,
\mathrm{d}t = -E_n T/{\hbar} $
is the dynamical phase and
$ \gamma_n(\mathcal{C}) \equiv \mathrm{i} \int_\mathcal{C}
\left\langle \Psi_n(\mathbf{R})| \partial_\mathbf{R}
\Psi_n(\mathbf{R}) \right\rangle  \cdot \mathrm{d}\mathbf{R}-
\alpha^n_{\mathbf{R}(T)} $
is the Berry phase (independent of $T$ and depending only on the
circuit $\mathcal{C}$), of interest here. Note that the last term,
in the above equation, is due to the multivaluedness of the basis
functions and was absent in the original paper of Berry
\cite{Berry1984}, where a single-valued basis was considered.

Simple algebra then yields
$ \gamma_n(\mathcal{C}) = \int_\mathcal{C} \mathbf{A}^n_\mathbf{R}
\cdot \mathrm{d}\mathbf{R} -\mathrm{i} \ln\left[\left\langle
\Psi_0^n \right| U_{\mathbf{R}(T)} \left| \Psi_0^n \right\rangle
\right] , $
with
$ \mathbf{A}^n_\mathbf{R} \equiv \left\langle \Psi^n_0\right|
\mathbf{A}_\mathbf{R} \left| \Psi^n_0 \right\rangle , $
and
$ \mathbf{A}_\mathbf{R} \equiv \mathrm{i} \left( U_\mathbf{R}
\right)^{-1} \left( \partial_\mathbf{R} U_\mathbf{R} \right) .$
One then obtains the components of $\mathbf{A}$:
$ A_\varphi = J_z (\cos\theta -1) - \sin\theta \left[
J_x\cos(\beta -\varphi ) -J_y \sin (\beta - \varphi ) \right]$,
$A_\theta = J_x \sin (\beta - \varphi ) + J_y \cos (\beta -
\varphi )$, $A_\beta = J_z$.
So far, we have not specified any particular choice for the
cartesian axes. When calculating $\gamma_n(\mathcal{C})$, it is
convenient to choose the $z$-axis along the expectation value of
$\mathbf{J}$ for the state $\left|\Psi^n_0\right\rangle$, i.e.,
$\hat{\mathbf{z}}_n \equiv {\mathbf{J}_n}/{J_n}$, with
$\mathbf{J}_n \equiv
\left\langle\Psi^n_0\right|\mathbf{J}\left|\Psi^n_0\right\rangle$
and $J_n \equiv \left\|\mathbf{J}_n \right\|$. Note that in the
most general situation, $J_n$ is not a multiple of $1/2$ and that
the $z$-axes defined in this way may be different for different
states $n$ and $n^\prime$; if $J_n=0$, the choice of
$\hat{\mathbf{z}}_n$ is indifferent. With this choice, one
immediately gets
$ A_\varphi^n = J_n \cos(\theta_n -1) $, 
$A_\theta^n = 0 $, 
$A_\beta^n = J_n , $
where the $n$ indices on the Euler angles remind that they are
defined here with respect to a $n$-dependent $z$-axis
$\hat{\mathbf{z}}_n$.

We now consider the last term of the Berry phase. Since $
\mathbf{R}(T)$ belongs to the symmetry group of $H_0$, it must
leave $\mathbf{J}_n$ invariant, so that $\theta_n(T) =0$;
therefore $U_{ \mathbf{R}(T)} = \mathrm{e}^{-\mathrm{i}\beta_n (
\mathcal{C})J_z}$, where
$\beta_n ( \mathcal{C}) \equiv \int_\mathcal{C} \mathrm{d}\beta_n$
is the total twist angle of the $x$- and $y$-axes around
$\hat{\mathbf{z}}_n$. If $\hat{\mathbf{z}}_n$ is a symmetry axis
of order $q$, we must have $\beta_n ( \mathcal{C}) =
p_n(\mathcal{C})\, 2\pi /q$, with $p_n(\mathcal{C})\in
\mathbb{Z}$.
The state
$|\Psi_0^n \rangle$ may be expanded in terms of the eigenstates
$|M;\hat{\mathbf{z}}_n \rangle$ of $J_z$ with quantum number $M$
(for the $z$-axis along $\hat{\mathbf{z}}_n$), i.e.,
$ |\Psi_0^n \rangle \equiv \sum_M a_M^n |M;\hat{\mathbf{z}}_n
\rangle . $
Let $M_n$ be the largest value of $M$ for which $a_M^n \neq 0$;
one can easily see that the only values of $M$ for which $a_M^n
\neq 0$ are of the form $M_n -rq$, with $r\in \mathbb{N}$, so that
$\alpha^n_{\mathbf{R}(T)}=M_n\,\beta_n(\mathcal{C}) \
\mathrm{mod}\ 2\pi$.
Putting everything together, one finally obtains
\begin{equation}\label{eq_final}
\gamma_n( \mathcal{C})\! = - J_n\, \Omega_n ( \mathcal{C} )
-\left( M_n \! -\! J_n \right) \frac{2\pi}{q}\, p_n( \mathcal{C})
\ \, \mathrm{mod}\, 2\pi ,
\end{equation}
where
$ \Omega_n( \mathcal{C}) \equiv \int_\mathcal{C}\left(1-
\cos\theta_n \right) \, \mathrm{d}\varphi_n $
is the (oriented) solid angle of the curve described by
$\hat{\mathbf{z}}_n( \mathbf{R})$. Equation (\ref{eq_final}) is
the central result of this paper. It is in fact a fairly general
theorem, applying a broad class of systems, as it relies only on
the properties of rotations, independently of any specific detail
of the Hamiltonian.

\begin{figure}[t]
\includegraphics[width=1.00\columnwidth]{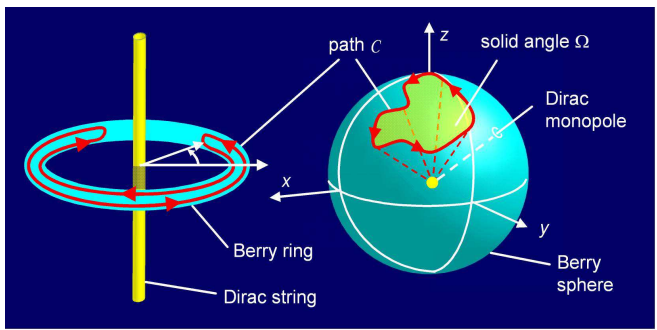}
\vspace*{0.1cm} \caption{Schematic representation of a circuit
$\mathcal{C}$ on the Berry sphere with a Dirac monopole at its
center, and on the Berry ring threaded by a Dirac string. In the
case depicted here, the winding number $p_n( \mathcal{C})$ equals
zero.} \label{fig_Dirac}
\end{figure}

The physical interpretation is as follows. A given rotation
$\mathbf{R}$ is represented by a point of polar coordinates
$(\varphi_n,\theta_n)$ on a sphere (hereafter called the Berry
sphere) giving the orientation of $\hat{\mathbf{z}}_n(
\mathbf{R})$, \emph{and} by a point on a ring (hereafter called
the Berry ring), with angular coordinate $q \beta_n$ (describing
the twist of the $x$- and $y$-axes around $\hat{\mathbf{z}}_n(
\mathbf{R})$); the winding number $p_n (\mathcal{C})$ of the
circuit $\mathcal{C}$ around the Dirac string is a topological
invariant of $\mathcal{C}$ (mod $2q$). A Dirac monopole of
strength $-J_n$ is positioned at the center of the sphere, whereas
the ring is threaded by a Dirac string carrying a flux equal to
$-2\pi (M_n - J_n )/q$. The Berry phase for a given circuit
$\mathcal{C}$ is then given by the sum of the fluxes through
$\mathcal{C}$ due to the Dirac monopole at the center of the Berry
sphere and to the Dirac string threading the Berry ring, as
depicted schematically on Fig.~\ref{fig_Dirac}.

The contribution of the Dirac monopole is proportional to the
solid angle $\Omega_n( \mathcal{C})$, which is a \emph{geometric}
property of the circuit $\mathcal{C}$, and may be called the
\emph{geometric} Berry phase. The contribution of the Dirac string
is proportional to the winding number $p_n(\mathcal{C})$ of the
circuit $\mathcal{C}$ around the Dirac string, and may be called
the \emph{topological} Berry phase.

The latter contribution constitutes a spin analogue of the
Aharonov-Bohm effect \cite{Aharonov1959}. The analogy, however is
not one-to-one, since the topology of the parameter space in the
present case ($\mathcal{M}=\mathbb{R}P^3/\mathbb{Z}_q$,
$\pi_1(M)=Z_{2q}$) differs from the one of the Aharonov-Bohm
effect ($\mathcal{M}=S^1 $, $\pi_1(M)=Z$). This is related to the
fact that a solid angle is defined modulo $4\pi$: writing
$p_n(\mathcal{C})=2qk_n(\mathcal{C})+r_n(\mathcal{C})$ with
$r_n(\mathcal{C})$ and $k_n(\mathcal{C})$ integers,
Eq.~(\ref{eq_final}) may be rewritten as
$\gamma_n( \mathcal{C}) = - J_n\, \left[ \Omega_n ( \mathcal{C} )
-4\pi k_n(\mathcal{C})\right] -\left( M_n \! -\! J_n \right)
\frac{2\pi}{q}\, r_n( \mathcal{C}) \ \, \mathrm{mod}\ 2\pi$.

The Dirac monopole giving rise to the geometric Berry phase is
generally non-quantized; the deviation from exact quantization,
$M_n-J_n$, is a measure of the effect of anisotropy for
$|\Psi_0^n\rangle$. This clearly illustrates the above discussion
on the interplay of charge quantization and spacial isotropy. The
\emph{topological} Berry phase due to the Dirac string obtained
here constitutes a novel effect that appears only for anisotropic
systems.

\begin{figure}[t]
\begin{center}
\includegraphics[width=0.80\columnwidth]{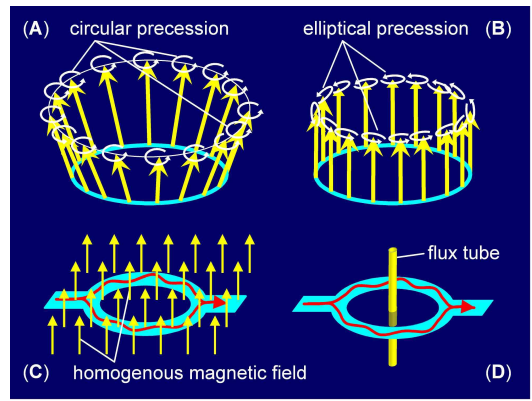}
\end{center}
\caption{Illustration of the analogy between (A) the
\emph{geometric} Berry phase of magnons on a ring due to a texture
of the magnetization and (C) the (\emph{geometric}) Aharonov-Bohm
effect in a homogenous field, contrasted with the analogy between
(B) the \emph{topological} Berry phase of magnons on a uniformly
magnetized ring due to the magnetic anisotropy and (D) the ``true"
(\emph{topological}) Aharonov-Bohm effect due to the vector
potential of a flux tube.} \label{fig_Magnon}
\end{figure}

Let us now consider how the \emph{geometric} and
\emph{topological} Berry phases could be probed experimentally.
The present theory most directly applies to nuclear or electronic
spin systems in an anisotropic environment. The simplest way of
probing the Berry is to repeat periodically the circuit
$\mathcal{C}$ at a frequency $\omega_\mathcal{C}/(2\pi)$ (with
$\hbar\omega_\mathcal{C}\ll |E_m - E_n|$, $\forall\ m\neq n$). The
Berry phase will therefore increase linearly in time, which means
that the energy of the state $n$ will be shifted by an amount
$\Delta E_n =-\hbar\omega_\mathcal{C} \, \gamma_n( \mathcal{C})
/(2\pi )$. These frequency shifts would be measurable as a shift
$\Delta\omega = \omega_\mathcal{C} \,( \gamma_n( \mathcal{C}) -
\gamma_{n^\prime}( \mathcal{C}))/(2\pi )$ of the magnetic
resonance line corresponding to transitions between the states $n$
and $n^\prime$. By taking suitably chosen circuits $\mathcal{C}$
one can investigate the geometric and adiabatic Berry phases
separately.

In particular, by choosing a circuit that is just a rotation of
$2\pi /q$ around $\hat{\mathbf{z}}_n$ (which is also the simplest
experiment), only the \emph{topological} Berry phase is probed.
For nuclear spins, this is simply achieved by spinning the sample,
a standard technique in nuclear magnetic resonance that was
successfully applied to study both the abelian and non-abelian
(\emph{geometric}) Berry phases of ${}^{35}\!$Cl nuclei ($S=3/2$)
\cite{Tycko1987, Zwanziger1990}. The simplest case is for a single
spin $S \geq 1$ with Hamiltonian $H_0=-BS_z + K{S_x}^2$ (case (b)
on Fig.~1). Taking $\mathcal{C}$ as a rotation of $+\pi$ around
the $z$-axis, one easily calculates the Berry phase: $\gamma_1 =
-\gamma_{-1}= - \pi [ 1- b/\sqrt{1+b^2}]$, $\gamma_0=0$, for
$S=1$, and $\gamma_{3/2}=-\gamma_{-1/2}=-\pi
[1-(b+1)/\sqrt{(b+1)^2+3}]$, $\gamma_{1/2}=-\gamma_{-3/2}=-\pi
[1-(b-1)/\sqrt{(b-1)^2+3}]$, for $S=3/2$, where $b\equiv 2B/K$ and
where the states are labelled by the (quantized) value of $S_z$
obtained in the limit $K\rightarrow 0$.

For electronic spins, sample spinning is not a realistic approach.
The Berry phase can nevertheless be investigated by applying the
magnetic field at some angle $\theta_B$ with respect to the
anisotropy axis (case (a) in Fig.~1), and by rotating the field on
a cone of angle $\theta_B$ around the anisotropy axis, which can
be conveniently done experimentally. The Berry phase obtained in
that case contains both a \emph{geometric} and a
\emph{topological} contribution.

\begin{figure}[t]
\begin{center}
\includegraphics[width=\columnwidth]{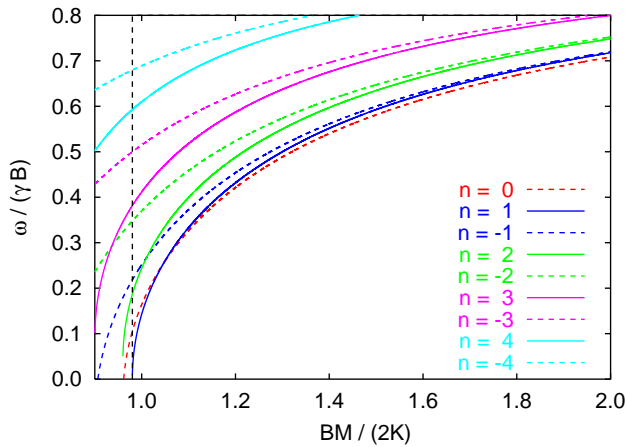}
\end{center}
\caption{Magnon spectrum of a ferromagnetic ring of radius $R$,
with exchange stiffness $A$ and tangential easy-axis anisotropy
$K=\pi M^2$, in a uniform perpendicular magnetic field $B$, for a
typical value $4KR^2/A=200$. Solid and dashed lines correspond to
modes with counterclockwise and clockwise group velocities,
respectively. The vertical dashed line gives the critical field
for stability of the state with uniform vertical magnetization.}
\label{fig_Spectrum}
\end{figure}

The Berry phase may also influence the properties of magnons. For
example, it has been recently indicated that the \emph{geometric}
Berry phase due to a non-coplanar texture of the magnetization of
a ferromagnetic ring (see Fig.~\ref{fig_Magnon}A) would affect the
dispersion of magnons (lifting the degeneracy of clockwise and
anticlockwise propagating magnons), and generate some equilibrium
spin currents \cite{Schutz2003}. In this study, however, the
effect of magnetic anisotropy was ignored (in a classical picture,
this correspond to a circular precession of the magnetization), so
that the resulting Berry phase associated with the propagation of
a magnon around the ring is merely that of a quantized monopole,
and does not include the \emph{topological} Berry phase due to a
non-quantized Dirac string. The \emph{geometric} Berry phase
obtained in this case is analogous to the \emph{geometric} (i.e.,
dependent on the particular geometry of electron trajectories,
because of the finite width of the ring's arms) Aharonov-Bohm
effect for a ring in a homogenous magnetic field, as depicted
schematically on Fig.~\ref{fig_Magnon}C. Since magnons have a spin
$S=1$, they may be subject to magnetic anisotropy. If one properly
incorporates the effect of magnetic anisotropy, then the
\emph{topological} Berry phase gives rise to new effects. For
example, if one considers a magnetic ring \emph{uniformly
magnetized} along the ring axis, and with some magnetic anisotropy
giving a tangentially oriented easy-magnetization axis, then the
precession of the magnetization (in a classical picture) has an
elliptical polarization whose large axis is tangential to the ring
and makes a turn of $2\pi$ around the ring (this corresponds to a
winding number $p(\mathcal{C})=2$), as depicted schematically on
Fig.~\ref{fig_Magnon}B, so that a topological Berry phase is
generated by the anisotropy; the degree of ellipticity, and hence
the topological Berry phase, can be conveniently controlled by
means of an external field parallel to the ring axis. The latter
situation is analogous (except for the difference discussed
earlier) to the ``true" (i.e., \emph{topological}) Aharonov-Bohm
effect, in which the flux threading the ring is concentrated in a
flux tube, the ring itself being in a field-free region (see
Fig.~\ref{fig_Magnon}D). The required anisotropy would be easily
obtained as a consequence of shape (dipolar) anisotropy, if one
takes a magnetic ring approximately as thick as wide. The
topological Berry phase would manifest as a splitting of magnon
spectrum, lifting the degeneracy between clockwise and
anticlockwise propagating magnons, an effect that could be
observed rather easily. A detailed account of the effect of Berry
phase on magnons will be given elsewhere \cite{Dugaev2004}; I give
below, without further details, the results for the magnon
spectrum of a ring of radius $R$ with exchange stiffness $A$ and
(dipolar) magnetic anisotropy $K=\pi M^2$, in a perpendicular
field $B$ sufficiently large to homogenously magnetize the ring
along its axis (see Fig.~\ref{fig_Magnon}B). I consider a typical
ferromagnetic ring (Ni ring of radius $R\simeq 75$~nm, width
$w\simeq 20$~nm, and thickness $h\simeq 20$~nm); the ring is
characterized by the dimensionless parameter $4KR^2/A \simeq 200$.
The calculated magnon spectrum is shown on
Fig.~\ref{fig_Spectrum}, where the lifting of the degeneracy
between states with anticlockwise ($n>0$) and clockwise ($n<0$)
group velocity due to the topological Berry phase appears very
clearly.

I am grateful to V.K.~Dugaev, C.~Lacroix, and B.~Canals for
intensive discussions throughout the elaboration process of this
work. This work was partly supported by the BMBF (Grant
No.~01BM924).

\vspace*{-\baselineskip}


\begin{thebibliography}{10}

\bibitem{Dirac1931}
P.A.M.~Dirac, Proc. Roy. Soc. London A \textbf{133}, 60 (1931);
Phys. Rev. \textbf{74}, 817 (1948).

\bibitem{Aharonov1959}
Y.~Aharonov and D.~Bohm, Phys. Rev. \textbf{115}, 485 (1959).

\bibitem{Baumann1988}
J.~Baumann \emph{et al.}, Phys. Rev. D \textbf{37}, 3107 (1988).

\bibitem{Berry1984}
M.V.~Berry, Proc. R. Soc. London A \textbf{392}, 45 (1984).

\bibitem{Tycko1987}
R.~Tycko, Phys. Rev. Lett. \textbf{58}, 2281 (1987).

\bibitem{Zwanziger1990}
J.W.~Zwanziger \emph{et al.}, Phys. Rev. A \textbf{42}, 3107
(1990).

\bibitem{Schutz2003}
F.~Sch{\"u}tz \emph{et al.}, Phys. Rev. Lett. \textbf{91}, 017205
(2003).

\bibitem{Dugaev2004}
V.K.~Dugaev, P.~Bruno, \emph{et al.}, to be published.

\end{thebibliography}
\end{document}